\documentstyle[11pt,aaspp4]{article}

\def\kms{km~s$^{-1}$}

\lefthead{Small et al.}
\righthead{Temporal Changes in QSO Broad Lines}

\begin{document}

\title{Temporal Changes in Quasar Broad Emission Line Profiles and the
Gravitationally Lensed Nature of Q1634+267A,B and Q2345+007A,B}

\author{Todd A. Small\altaffilmark{1}, Wallace L.W. Sargent,
 and Charles C. Steidel}
\affil{Palomar Observatory, California Institute of Technology,
Pasadena, CA 91125 \\
Electronic mail: tas@ast.cam.ac.uk; wws,ccs@astro.caltech.edu}

\altaffiltext{1}{present address:  Institute of Astronomy, 
University of Cambridge, Madingley Road, Cambridge CB3 0HA, United Kingdom}

\begin{abstract}

Steidel \& Sargent (1991a) \nocite{Steidel:Q1634} found that the
spectra of the components of the close quasar pairs Q1634+267A,B and
Q2345+007A,B were in each case remarkably similar (except for overall
differences in brightness) but that individual broad emission lines
due to Ly $\alpha$, \ion{Si}{4}, \ion{C}{4} and \ion{C}{3}] exhibited
differences in profile or equivalent width between the A and B
components. If the A and B components of these objects result from
gravitational lensing of a single QSO, the difference in light travel
time for the two images is roughly one year for both pairs.
Accordingly, the authors suggested that observations of the spectra of
single QSOs would reveal similar changes in emission line profiles and
equivalent widths on a time scale of one year or less.  In order to
test Steidel \& Sargent's (1991a) \nocite{Steidel:Q1634} hypothesis,
we obtained in 1992 high quality spectra of 30 quasars with $z_{em}
\sim 1$ selected from the brighter QSOs in the sample of Steidel \&
Sargent (1991b)\nocite{Steidel:emlines}, which those authors had
observed in 1989--90. The new spectra were obtained with exactly the
same instrumental setup as the earlier observations. The two sets of
spectra were reduced in an identical manner and then compared in order
to search for changes in the strengths and profiles of the principal
emission lines occurring on time scales of order one year.  We find
that the spectra from the two epochs are remarkably similar for all
the QS0s, except for overall changes in fluxes and minor differences
in continuum shape which are probably artifacts introduced by the
observing procedures. However, there are also changes in the
equivalent widths or shapes of the stronger emission lines on
rest-frame time scales of 1-1.5 years in about two thirds of the
QSOs.  These observations, therefore, support the notion that
Q1634+267A,B and Q2345+007A,B are gravitationally lensed and that the
spectra of the components in lensed QSOs need not be exactly
identical. Moreover, it should be expected that any differences
between the components in such systems should change with time.
  
\end{abstract}

\section{Introduction}

The nature of two close pairs of QSOs, Q1634+267A,B (separation
3.77\arcsec) and Q2345+007A,B (7.03\arcsec) is still not established
with certainty. In both cases, the two components of the pair have the
same redshift within the errors of measurement and, as a result, they
have both been considered likely to result from gravitational lensing
of a single QSO (\cite{Djorgovski:Q1634}, \cite{Weedman:Q2345}).
However, a candidate for the lens has not been established with
certainty in either case, despite the fact that such wide spacings
require either an unusually massive galaxy or a cluster of galaxies.
Deep infrared observations at the Keck 10-meter telescope of
Q1634+267A,B have not revealed a candidate lensing galaxy down to a
3.5 $\sigma$ detection limit of K = 21.3 mag, although we note as well
that the object responsible for the strong absorption line system at
$z \approx 1.126$ was also not detected (\cite{Neugebauer:Q1634},
J. Larkin, personal communication).  A $B_{\rm J} = 25.0$ lens
candidate for Q2345+007A,B has recently been detected by Fischer et
al. (1994) \nocite{Fischer:Q2345}northeast of image B, but it will
take at least a 10-meter-class telescope to measure the redshift of
the candidate.  Bonnet et al.\ (1993) \nocite{Bonnet:Q2345} have both
reported arclets and lensing-induced distortions of faint galaxies
near the pair, which is further evidence for the presence of a massive
lens.  Using redshifts of galaxies in the field of Q2345+007A,B
estimated with multi-color photometry, Pell\'{o} et al.\ (1996)
\nocite{Pello:Q2345} have found evidence for a cluster of galaxies at
$z \approx 0.78$, but the cluster core lies 45\arcsec\ away from the
lensed pair.  Accordingly, the possibility is still entertained that
the pairs are binary QSOs with separations of a few tens of kpc. A
resolution of the question is of considerable importance because such
QSO pairs are essential for measuring the sizes of the intervening
absorption systems (see, e.g., \cite{Foltz:ly_alpha};
\cite{Duncan:ly_alpha}; \cite{McGill:ly_alpha}; \cite{Steidel:Q1634};
\cite{Bechtold:ly_alpha}; Dinshaw et al.\ 1994,
1995\nocite{Dinshaw:ly_alpha1}\nocite{Dinshaw:ly_alpha2}); however,
the limits determined for the sizes depend strongly on whether a
particular close pair is a manifestation of lensing or a real binary
QSO.

Steidel \& Sargent (1991a) \nocite{Steidel:Q1634} recorded high
signal-to-noise ratio spectra of the QSO pairs Q1634+267A,B and
Q2345+007A,B in order to examine critically the degree of similarity
of the two components in each pair.  The two components were observed
{\it simultaneously} with a long slit.  They found that the overall
continuum shapes in the spectra of each pair were remarkably similar
and that, in general, the emission lines closely matched.  However,
they also discovered that the Ly $\alpha$, \ion{N}{5}, \ion{C}{4}, and
\ion{Si}{4} emission lines of the two images of Q1634+267A,B exhibit
differences in profile, which they described as velocity shifts of as
much as $\sim$ 1000 \kms. On the other hand, while the redshifts of
the emission lines in the spectra of Q2345+007A,B are
indistinguishable within the measurement errors, ($\Delta v_{B-A} = 15
\pm 20$ \kms), the Ly $\alpha$ and, particularly, the \ion{C}{3}]
emission lines differ significantly in strength in the two spectra.
Because of the remarkable overall similarity of the spectra of the
components within each pair, Steidel \& Sargent (1991a)
\nocite{Steidel:Q1634} concluded that both Q1634+267A,B and
Q2345+007A,B result from gravitational lensing. Since the prominent
spectral differences are too large to be comfortably accommodated as
microlensing by stars in the lens (\cite{Nemiroff:microlensing},
\cite{Schneider:microlensing}), they ascribed the differences to
intrinsic emission line profile variations on a characteristic
time scale less than the time delays between the light travel times of
the two images, which are roughly one year for both pairs.

Variability in the strengths and profiles of quasar broad emission
lines on year and shorter time scales has been reported by a number of
workers (e.g., \cite{Zheng:emlines}; \cite{O'Brien:emlines}; and
\cite{Gondhalekar:emlines}), but these reports have been based on
either observations of the Balmer lines or on comparatively noisy
spectra of the high-ionization ultraviolet lines obtained with {\it
International Ultraviolet Explorer} satellite.  Following Steidel \&
Sargent's (1991a) \nocite{Steidel:Q1634} work, Corbin (1992)
\nocite{Corbin:emlines} announced that he had discovered increases in
the equivalent widths acccompanied by shifts of 2200 \kms\ in the
redshifts of the \ion{C}{3}] emission line in the spectra of two
QSOs. The spectra had been observed 1.5 years apart.  This work has
not been fully published or critically assessed and, even if it proves
to be correct, there is no information on the ubiquity of such
behavior, although there have been other reports of profile
differences in the broad lines of candidate gravitationally-lensed
quasars pairs (e.g., \cite{Filippenko:mg2}, \cite{Wisotzki:he1104},
\cite{Impey:B1422}).  In particular, there are two reports of dramatic
velocity differences similar to those described by Steidel \& Sargent
(1991a) \nocite{Steidel:Q1634}.  Saunders et al.\ (1997)
\nocite{Saunders:pc1643} argue that the quasar pair PC 1643+4631A,B
($z = 3.79, 3.83$, separation 198\arcsec) is lensed by a cluster
responsible for the cosmic microwave background decrement detected by
Jones et al.\ (1997) \nocite{Jones:pc1643} in the direction of the
quasar pair.  Michalitsianos et al.\ (1997)
\nocite{Michalitsianos:um425} describe a velocity difference of $\sim
600-700$ \kms\ between the broad emission lines of UM425A,B, a pair of
$z \approx 1.47$ quasars with a 6.5\arcsec\ separation.

There is a persuasive indirect argument for believing that the
centroid velocities of the broad emission lines in QSOs must vary.
Namely, it is observed that the velocities determined from different
ions in a quasar spectrum often differ by $\sim$ 1000 \kms.  Since the
distribution of velocity differences is broad with a mode that is
consistent with 0 \kms\ (\cite{Steidel:emlines}), it is likely that
these velocity differences reflect variability in the velocities of
the broad lines, although on an unknown time scale.

Based on the above considerations, we decided to search for emission
line profile and velocity changes in single QSOs by reobserving 30
objects from the \ion{Mg}{2} survey of Steidel \& Sargent (1992)
\nocite{Steidel:mg2}, using an identical instrument configuration. The
spectra used in the survey were obtained for measuring {\it
absorption} lines, in particular, the \ion{Mg}{2}
$\lambda\lambda$2796,2803 doublet; accordingly, their S/N ratio is
exceptionally high. In order to study the cosmological evolution of
\ion{Mg}{2} absorption, the QSOs were selected to have redshifts in
two ranges, one around $z_{em} \sim 1.2$ and the other around $z_{em}
\sim 2$. In general, the spectra cover the wavelength region from
below the \ion{C}{3}] $\lambda 1909$ emission line to above the
\ion{Mg}{2} $\lambda 2800$ emission line. We chose to repeat only
spectra of QSOs in the lower redshift range because the sky
subtraction is much harder for QSOs in the higher redshift range. We
also elected to reobserve the brighter QSOs in the original survey
where possible. The observing and data reduction procedures are
described in \S\ 2, an analysis of the spectral differences between the
two epochs is given in \S\ 3, and we discuss our conclusions in \S\ 4.

\section{Observations and Data Reduction}

We have reobserved 30 quasars with $z \sim 1$ from the survey of
Steidel \& Sargent (1992)\nocite{Steidel:mg2}.  We used an identical
instrumental setup: the Double Spectrograph attached to the 5.08m Hale
Telescope, covering the spectral range from 3100 - 7000\AA\ with $\sim
4$\AA\ resolution for wavelengths $\lesssim 4700$ \AA\ (the blue side)
and $\sim 6$\AA\ resolution for wavelengths $\gtrsim 4700$ \AA\ (the
red side). The observations were made with a 1\arcsec\ slit. A journal
of our observations is presented in Table \ref{tables:journal}.  We
were careful to set the position angle of the slit to the parallactic
angle appropriate to the midpoint of the observation in order to
minimize light losses at the blue end of the spectrum.  However,
because of seeing fluctuations and the narrow entrance slit, we can
say nothing about changes in continuum fluxes.  Moreover, the
possibility of guiding errors, together with the fact that the
position angle of the slit was not adjusted during the typically
hour-long exposures, means that apparent changes in the overall
continuum slopes between the two epochs cannot be trusted.

Our data were reduced using standard techniques and routines from
IRAF.  The spectra were flux-calibrated using instrumental sensitivity
functions determined from observations of spectrophotometric
standards.  The spectra usually matched at the overlap region between
the blue and red cameras.  We found that while our reductions of the
original data of Steidel \& Sargent (1992) \nocite{Steidel:mg2} agreed
very well in the red, there was significant disagreement in the blue.
This disagreement is due to the fact that Steidel \& Sargent (1992)
\nocite{Steidel:mg2} used the airmass at the end of the
exposure\footnote{Since Steidel \& Sargent (1992) \nocite{Steidel:mg2}
were searching for absorption lines, the errors caused by using the
airmass at the end of the exposure have no effect on their analysis.},
which is the airmass recorded in the image headers at Palomar, to
compute the extinction correction; whereas, we used the mean airmass
appropriate for the whole exposure.  The comparisons between the two
epochs are based exclusively on the re-reduced data.

\section{Analysis}

\subsection{Comparison of the Spectra}

As described above (\S\ 2), we do not trust apparent changes in the
continua between the two epochs, primarily because of our use of a
narrow slit.  Since in this study we are only interested in the
emission lines, we have chosen to remove the continuum variations by
fitting 5 piece cubic splines to the spectra.  The composite QSO
spectrum synthesized by Francis et al. (1991) \nocite{Francis:quasar}
from 718 individual spectra has many emission lines with few gaps
where continuum may be present.  Accordingly, our fits are not meant
to be accurate representations of the true quasar continua.  The fits
were arranged to ignore the broad emission lines and the narrow
absorption lines by using a rejection threshold of 2 $\sigma$ for
positive deviations and a threshold of 4 $\sigma$ for negative
deviations.  As can be seen in Figure \ref{figures:quasars}, the
continuum-flattened spectra from the two epochs match very well.
However, there is a distinct tendency for the blue sections of the
spectra to agree better than the red sections. This is surprising
because both guiding errors and the incomplete correction for
atmospheric dispersion would be expected to have a greater effect on
the repeatabilty of the blue ends of the spectra. However, the red
camera of the Double Spectrograph was prone to ice on the dewar window
which was hard to recognize during the course of observations until
the effects become very severe.

\subsection{Notes on Individual Objects}

We have compared the spectra of each QSO obtained at the two epochs in
two ways.  First we compared the blue and red spectra shown in Figure
\ref{figures:quasars} separately, then we compared the complete
spectra. For this latter purpose color renditions of the spectra
obtained at the two epochs were superimposed; these color pictures are
not reproduced here, but digital copies can be supplied to interested
readers on request to the first author. Our conclusions regarding
changes in the spectra of each object are given below.
 
{\it a) Q0024+2225; $z_{em} = 1.118$}.
The continuum-normalized spectra match very well except below
3200 \AA.  The prominent broad emission lines, \ion{C}{4}, \ion{C}{3}],
and \ion{Mg}{2}, are all stronger in the first epoch spectrum.  

{\it b) Q0117+2118; $z_{em} = 1.493$}. 
The spectra from the two epochs are in excellent agreement except for
the \ion{Si}{4} line, which is stronger in the first epoch spectrum.

{\it c) Q0232-0415; $z_{em} = 1.434$}.
While the red side spectra agree satisfactorily, the blue side spectra
agree quite poorly.  Both \ion{C}{4} and \ion{C}{3}] exhibit dramatic
changes in equivalent width. We believe that the the change in
\ion{C}{4} is real; the apparent change in \ion{C}{3}] could arise
from problems with the calibration of the spectra near the dichroic
changeover. The pair of emission lines at $\sim 4000$ \AA, He II
$\lambda 1640$ and O III] $\lambda 1663$, appear to be resolved in the
second epoch spectrum and merged together at the first epoch.  An
examination of the other spectra in our sample reveals that the
resolution of these features varies considerably from object to
object.

{\it d) Q0248+4302; $z_{em} = 1.316$ }.
The spectra match astonishingly well.
Spectra like these help give credence to the reality of the apparent 
changes in other spectra in the sample.  The broad emissions lines
are all stronger in the first epoch spectrum.

{\it e) Q0333+3208; $z_{em} = 1.263$}.
Both \ion{C}{3}] and \ion{Mg}{2} appear stronger in the second epoch
spectra.

{\it f) Q0454+0356; $z_{em} = 1.345$}.
The spectra from the two epochs match very well, and
there is notably good agreement between the many absorption lines.
The broad emission lines, particularly \ion{C}{3}], are stronger in the
first epoch spectrum.  In addition, \ion{C}{4} shows slight but
significant profile differences.

{\it g) Q0856+4649; $z_{em} = 0.924$}.
There are no believable changes.

{\it h) Q0859-1403; $z_{em} = 1.327$ }.
The spectra from the two epochs match well.  
The \ion{C}{4} and \ion{C}{3}] lines were stronger in the first epoch
spectrum.

{\it i) Q0946+3009; $z_{em} = 1.216$}.
This is a BAL QSO. The \ion{C}{4} and \ion{Mg}{2} emission lines were 
stronger in the first epoch spectrum. The red continua match particularly well.

{\it j) Q1019+3056; $z_{em} = 1.316$}.
The spectra from the two epochs agree very well except that the broad
emission lines are somewhat stronger in the first epoch spectrum.

{\it k) Q1038+0625; $z_{em} = 1.270$}.
The spectra from the two epochs match well.

{\it l) Q1120+0154; $z_{em} = 1.490$}.
The spectra from the two epochs match nicely; however, the complex BAL
absorption blueward of \ion{C}{4} appears to have changed.  We judge the
apparent changes in \ion{C}{3}] and \ion{Mg}{2} to be uncertain because the
lines fall in the dichroic crossover region and the extreme red end of
the spectra, respectively.  Neither \ion{Si}{4} nor
\ion{C}{4} show anys signs of variation.

{\it m) Q1206+4557; $z_{em} = 1.158$}.
The spectra agree nicely; however, the
\ion{C}{4}, \ion{C}{3}] and \ion{Mg}{2} emission lines are all
slightly stronger in the second epoch spectrum.

{\it n) Q1241+1737; $z_{em} = 1.273$}.
The spectra match well everywhere.
  
{\it o) Q1254+0443; $z_{em} = 1.024$}.
\ion{Mg}{2} in the second epoch spectrum is marginally stronger;
otherwise, the spectra agree well.  The \ion{C}{3}], \ion{Al}{3}
complex has an unusual profile at both epochs.

{\it p) Q1317+2743; $z_{em} = 1.022$}.
The spectra measured at the two epochs match satisfactorily. An unidentified 
feature observed at $\approx 4200$ \AA\ is stronger in the first epoch
spectrum, as are the red wing and center of the \ion{Mg}{2} emission line.

{\it q) Q1338+4138; $z_{em} = 1.219$}
Apart from poor agreement and excessive noise around the dichroic
changeover, the spectra from the two epochs match well. The
\ion{Mg}{2} and \ion{C}{4} emission lines show no changes, but the red
wing of the \ion{C}{3}] line is higher at the earlier epoch, as is an
unidentified feature at $\approx 5150$ \AA.  The variation of the \ion{C}{3}]
profile is reminiscent of the variation in \ion{N}{5} observed by
Steidel \& Sargent (1991a) \nocite{Steidel:Q1634} in Q1634+267A,B.

{\it r) Q1340+2859; $z_{em} = 0.905$}.
The spectra from the two epochs agree satisfactorily as does the
\ion{Mg}{2} profile.  However, \ion{C}{3}] is wider in the second epoch
spectrum.  A sharp [Ne V] $\lambda$3425 emission line is exactly the
same at each epoch.

{\it s) Q1356+5806; $z_{em} = 1.371$}.
The spectra match well.  The equivalent widths of the \ion{C}{4},
\ion{C}{3}] and \ion{Mg}{2} emission lines are greater in the first
epoch spectrum.

{\it t) Q1407+2632; $z_{em} = 0.944$}.
The complex spectra match very well.  The emission lines look very
unusual; the \ion{Fe}{2} lines are very strong, and \ion{C}{3}] and
\ion{Mg}{2} are remarkably weak.

{\it u) Q1458+7152; $z_{em} = 0.905$}.
The emission lines are unusually sharp; \ion{C}{3}] and \ion{Mg}{2}
are substantially stronger in the second epoch spectrum.  The [Ne V]
$\lambda$3425 emission line is stronger in the second epoch spectrum
as well.
 
{\it v) Q1522+1009; $z_{em} = 1.321$}.
The first epoch spectrum was recorded with the red camera grating
inadvertently adjusted to give a central wavelength $\sim 200$ \AA\ to
the red, and thus the continuum of the spectrum near the dichroic
changeover is quite uncertain.  Otherwise, the spectra agree well.
\ion{Mg}{2} is stronger and wider at the first epoch.  With less
certainty \ion{C}{4} is stronger at the second epoch.

{\it w) Q1538+4745; $z_{em} = 0.770$}.
The spectra and the \ion{Mg}{2} emission line profiles agree very
well. A  sharp [O II] $\lambda$3727 emission line is visible at
both epochs. The \ion{C}{3}] emission line is conspicuously wider in
the first epoch spectrum, due to a stronger blue wing.

{\it x) Q1634+7037; $z_{em} = 1.334$}.
The blue halves match very well.  Unfortunately,
Steidel \& Sargent did not take a usable standard star for the red camera on
the night that this object was observed during the first epoch, and
so the best we can do for flux calibration is to use a red camera
sensitivity function from another night. There is poor agreement near
the dichroic changeover. In view of the flaws in the data, it is difficult
to believe in the reality of the differences in the red spectra.
There is marginal evidence that \ion{C}{3}] is stronger in the first
epoch spectrum.

{\it y) Q1637+5726; $z_{em} = 0.745$}.
The spectra match unusually poorly, especially in the blue spectra.
\ion{Mg}{2} is noticeably stronger in the first epoch spectra.  \ion{C}{3}]
appears to be stronger, too, but the increased noise in the far blue
end of the spectrum makes this assessment less reliable.  The
narrow forbidden lines redward of \ion{Mg}{2}, especially [\ion{O}{2}],
agree nicely.

{\it z) Q1656+0519; $z_{em} = 0.887$}.
These complex spectra match well.

{\it aa) Q1718+4806; $z_{em} = 1.084$}.
The spectra from the two epochs agree comparatively poorly.  The broad
emission lines agree well, however.

{\it bb) Q1821+1042; $z_{em} = 1.360$}.
The spectra from the two epochs agree well.
The \ion{C}{4} line is identical, while \ion{C}{3}] and \ion{Mg}{2} are
slightly wider and stronger at the first epoch.

\begin{figure}
\plotfiddle{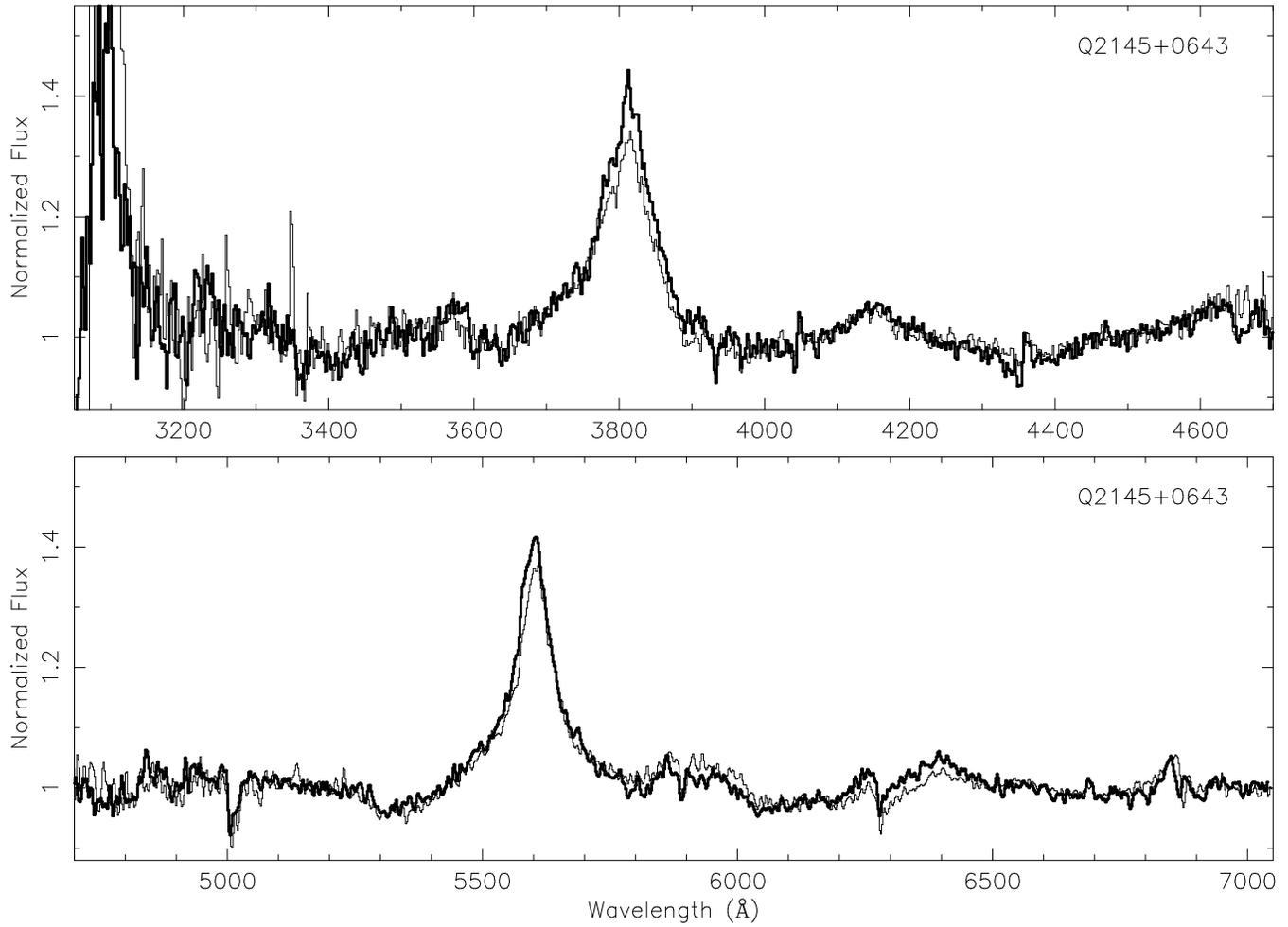}{4 in}{0}{100}{100}{-306}{-360}
\caption[]{
Continuum-flattened spectra of Q2145+0643.  The blue and red spectra
are displayed in separate panels.  The second epoch
spectrum is drawn with the heavier line.  The spectrum from 1992 October 20
is drawn with the lighter line.}
\label{figures:q2145}
\end{figure}

{\it cc) Q2145+0643; $z_{em} = 0.990$}.
The spectra from the two epochs agree well except near the dichroic
changeover.  \ion{Mg}{2} is slightly stronger in the first epoch
spectrum.  For another project, we obtained an additional spectrum of
this quasar on 1992 October 20 ($\sim 4$ months after the second epoch
spectrum) with the same instrument configuration.  In Figure
\ref{figures:q2145}, we plot the second epoch spectrum and the 1992
October 20 spectrum.  Despite the short time interval between the two
observations, we find significant changes in the strengths of
\ion{C}{3}] and \ion{Mg}{2}.
 
{\it dd) Q2216-0350; $z_{em} = 0.901$}.
The red side matches well; the blue side merely
satisfactorily. \ion{Mg}{2} is stronger at the second epoch. A weak, sharp
[Ne V] $\lambda$3425 emission line is identical in the two spectra.

\subsection{Cross-correlation analysis}

\begin{figure}
\plotfiddle{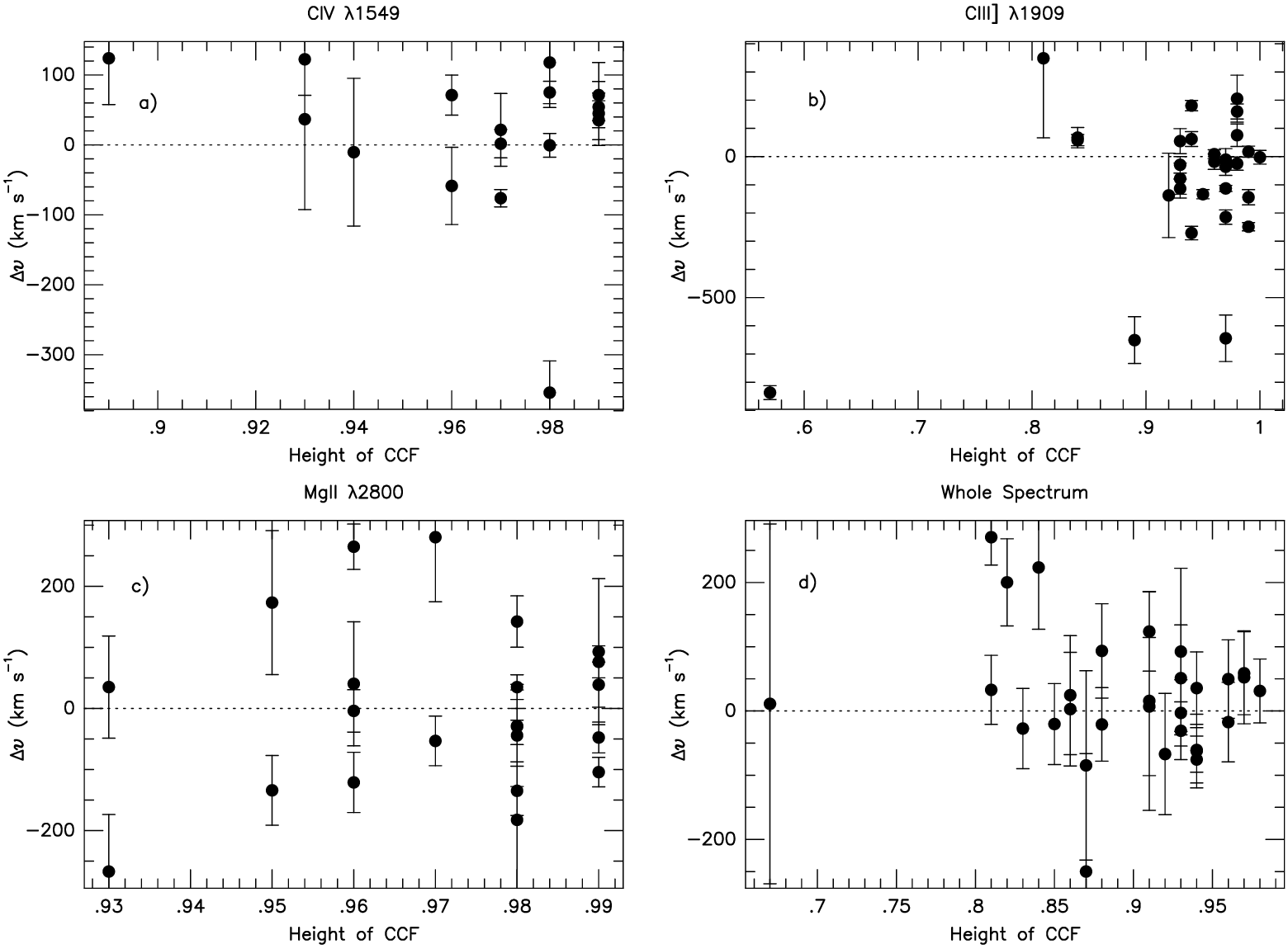}{5in}{0}{75}{75}{-306}{-36}
\caption[]
{Results of the cross-correlation of spectra from the two epochs.
The computed velocity difference is plotted against the cross-correlation peak
height for the (a) \ion{C}{4}, (b) \ion{C}{3}], and (c) \ion{Mg}{2}
emission lines and (d) for the whole observed wavelength range.  
The error bars show the formal errors from the cross-correlation analysis,
although we take the true errors to be the rms velocity difference of the
quasar pairs when whole wavelength range is correlated, 100 \kms.  Note
the different axis scales in each panel.}
\label{figures:xcor}
\end{figure}

Close visual inspection of the spectra reveals no significant emission line
velocity shifts; however, we have also performed a cross-correlation
analysis of the spectra to search systematically for such shifts.  The
results are summarized in Table 2 and in Figure
\ref{figures:xcor}.  We cross-correlated three emission lines,
\ion{C}{4}, \ion{C}{3}], and \ion{Mg}{2}, choosing sample regions
corresponding to the extent of the emission lines measured by Francis
et al.\ (1991) \nocite{Francis:quasar} in their composite quasar
spectrum.  The cross-correlation results are not terribly sensitive to
the precise wavelength interval chosen; increasing the windows by up
to 50\AA\ on both ends changes the measured velocity differences by
only $\approx$ 80 \kms.  The results are consistent with our
impressions from a visual examination of spectra, namely that there
are no significant shifts.  Although the formal errors from the
cross-correlation analysis (based on the formulae given in Tonry \&
Davis [1979] \nocite{Tonry:xcor}), are often only 10 or 20 \kms, we
took the true error to be the rms velocity difference of the quasar
pairs when the whole wavelength range from 3000 \AA\ to 7000 \AA\ was
correlated.  This value is 100 \kms.  (In cases in which the formal
error is actually larger than 100 \kms, we have simply used the formal
error.)  Using this estimate of the error, the cross-correlation
analysis finds four apparently significant shifts between the two
epochs, Q0454+0356 \ion{C}{4}, Q1120+0154 \ion{C}{3}], Q1407+2632
\ion{C}{3}], and Q1538+4745 \ion{C}{3}].  The velocity differences
found for Q0454+0356 \ion{C}{4} and Q1538+4745 \ion{C}{3}] agree with
our visual impressions.  The results for Q1120+0154 \ion{C}{3}] and
Q1407+2632 \ion{C}{3}] are not likely to be reliable because, in the
first case, the line lies on the dichroic changeover and, in the
second case, the line is unusually weak and broad.  Moreover, based on
our detailed examination of the spectra, we believe that a simple
velocity shift obtained from a cross-correlation analysis gives a poor
description of the complex emission line profile changes actually
observed.  Nevertheless, we note that we have computed the velocity
shifts of Ly$\alpha$ and \ion{C}{4} in Steidel \& Sargent's (1991a)
\nocite{Steidel:Q1634} original spectra of Q1634+267 by
cross-correlating the spectra of A and B and have found that the
shifts are $\sim$300-500 \kms, not $\sim$1000 \kms\ as these authors
reported.

\section{Discussion}

Many of the QSOs in our sample exhibit temporal variations in the
profiles of the broad emission lines similar to the observed
differences in line profile between the A and B images of Q1623+267
and Q2345+007.  These variations occur on time scales of 1-1.5 years
in the rest frames of the QSOs.  Accordingly, the relatively subtle
differences in the broad emission line profiles in these close pairs
could be due to the roughly one year difference in light travel time
between the A and B components. Consequently, our study supports
Steidel \& Sargent's (1991a) \nocite{Steidel:Q1634} conclusion that
Q1634+267A,B and Q2345+007A,B are lensed quasars and not physical
binary quasars.  In addition, our study highlights the fact that the
spectra of the components of a lensed quasar need not be identical and
that the differences should be expected to change with time.

In order to establish beyond doubt that Q1634+267A,B and Q2345+007A,B
are lensed systems, we suggest searching for correlated amplification
changes in the system on time scales of a year and making high
sensitivity radio observations of the system, similar to the one that
Djorgovski et al.\ (1987) \nocite{Djorgovski:Q1145} used to prove that
Q1145-071A,B was {\it not} a lensed system.  We note that Patnaik,
Schneider, \& Narayan (1996) \nocite{Patnaik:Q2345} have found weak
radio emission from Q1634+267A and the limit on the emission from the
B component is consistent with the optical brightness ratio.  The
approach of Bonnet et al.\ (1993) \nocite{Bonnet:Q2345} and Pell\'{o}
et al.\ (1996) \nocite{Pello:Q2345} of using weak gravitational lensing
to identify mass concentrations in the field Q2345+007A,B also appears
to be promising, but the possible massive cluster inferred from their
data requires spectroscopic confirmation.

\acknowledgments

We are grateful to the staff of the Palomar Observatory for making
these observations possible.  The comments of the referee, Craig Foltz,
helped us to improve the presentation.
This work has been supported in part by NSF Grant No. AST-92213165 (W.S.), an
NSF Graduate Fellowship (T.S.), and by NASA through Grant No. HF-1008.01-90A
awarded by the Space Telescope Science Institute which is operated by the
AURA, Inc. for NASA under Contract No. NAS5-26555 (C.S.).

\begin{deluxetable}{lcllc}
\tablecolumns{5}
\tablewidth{0pt}
\tablenum{1}
\tablecaption{Journal of Observations}
\tablehead{
\colhead{} &
\colhead{} & 
\multicolumn{2}{c}{Date (UT)} & 
\colhead{} \\ [.2ex]
\colhead{Quasar} &
\colhead{$z_{em}$} &
\colhead{1$^{\rm st}$ Epoch} &
\colhead{2$^{\rm nd}$ Epoch} &
\colhead{Type\tablenotemark{a}}}
\startdata	
Q0024$+$2225 & 1.118 & 1989 Jul 13 & 1992 Jan 31 & R \nl
Q0117$+$2118 & 1.493 & 1989 Nov 25 & 1992 Jan 31 & U \nl
Q0232$-$0415 & 1.434 & 1989 Nov 25 & 1992 Jan 31 & R \nl
Q0248$+$4302 & 1.316 & 1989 Nov 25 & 1992 Jan 31 & R \nl
       	   &       & 1989 Nov 27 &  	       &  \nl
	   &	   & 1990 Jan 20 & 	       &  \nl
Q0333$+$3208 & 1.263 & 1989 Nov 25 & 1992 Jan 31 & R \nl
           &       & 1990 Jan 21 &             &  \nl
Q0454$+$0356 & 1.345 & 1989 Nov 25 & 1992 Jan 31 & R \nl
	   &       & 1990 Jan 20 &             &  \nl
Q0856$+$4649 & 0.924 & 1990 Jan 22 & 1992 Jan 30 & U \nl
Q0859$-$1403 & 1.327 & 1990 Jan 22 & 1992 Jan 31 & Rs \nl
Q0946$+$3009 & 1.216 & 1990 Jan 21 & 1992 Jan 30 & U \nl
Q1019$+$3056 & 1.316 & 1990 Jan 22 & 1992 Jan 31 & Rs \nl
Q1038$+$0625 & 1.270 & 1990 Jan 21 & 1992 Jan 31 & Rs \nl
Q1120$+$0154 & 1.490 & 1990 Jan 22 & 1992 Jan 30 & S \nl
Q1206$+$4557 & 1.158 & 1989 Jul 14 & 1992 Jan 31 & U \nl
Q1241$+$1737 & 1.273 & 1989 Jul 13 & 1992 Jan 31 & U \nl
Q1254$+$0443 & 1.024 & 1990 Jan 20 & 1992 Jan 31 & U \nl
Q1317$+$2743 & 1.022 & 1989 Jul 12 & 1992 Jan 31 & R \nl
Q1338$+$4138 & 1.219 & 1989 Jul 13 & 1992 Jan 31 & U \nl
Q1340$+$2859 & 0.905 & 1989 Jul 14 & 1992 Jun 26 & R \nl
Q1356$+$5806 & 1.371 & 1989 Jul 13 & 1992 Jun 26 & R \nl
Q1407$+$2632 & 0.944 & 1989 Jul 13 & 1992 Jan 30 & Rs \nl
Q1458$+$7152 & 0.905 & 1989 Jul 12 & 1992 Jun 26 & Rs \nl
Q1522$+$1009 & 1.321 & 1989 Jul 11 & 1992 Jun 26 & U \nl
Q1538$+$4745 & 0.770 & 1989 Jul 14 & 1992 Jun 26 & U \nl
Q1634$+$7037 & 1.334 & 1989 Jul 11 & 1992 Jun 25 & U \nl
Q1637$+$5726 & 0.745 & 1989 Jul 13 & 1992 Jun 26 & R \nl
Q1656$+$0519 & 0.887 & 1989 Jul 11 & 1992 Jun 25 & R \nl
Q1718$+$4807 & 1.084 & 1989 Jul 11 & 1992 Jun 25 & U \nl
Q1821$+$1042 & 1.360 & 1989 Jul 11 & 1992 Jun 25 & R \nl
Q2145$+$0643 & 0.990 & 1989 Jul 11 & 1992 Jun 26 & R \nl
Q2216$-$0350 & 0.901 & 1989 Jul 14 & 1992 Jun 26 & R \nl
\enddata
\tablenotetext{a}{Method of original discovery: ``R'' denotes radio
selected, where ``Rs'' denotes ``steep'' radio spectrum ($\alpha > 0.5$, 
$f_\nu \propto \nu^{-\alpha}$), ``U'' denotes ultraviolet excess, and 
``S'' denotes slitless spectroscopy.}
\label{tables:journal}
\end{deluxetable}

\newpage
\clearpage

\begin{table}
\dummytable\label{tables:xcor}
\end{table}

\begin{figure}
\plotfiddle{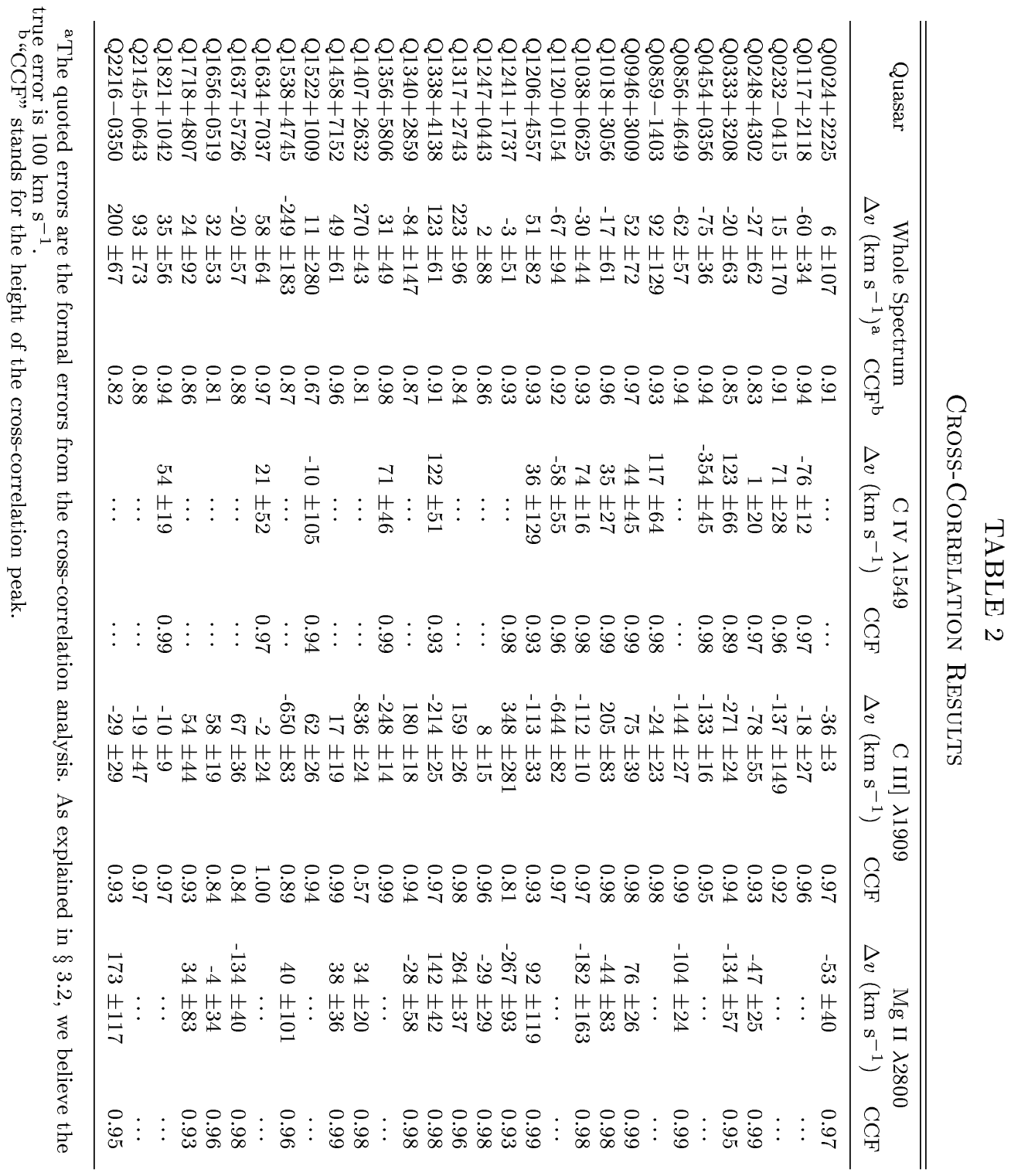}{5 in}{90}{100}{100}{427}{0}
\end{figure}

\newpage
\clearpage

\begin{figure}
\plotfiddle{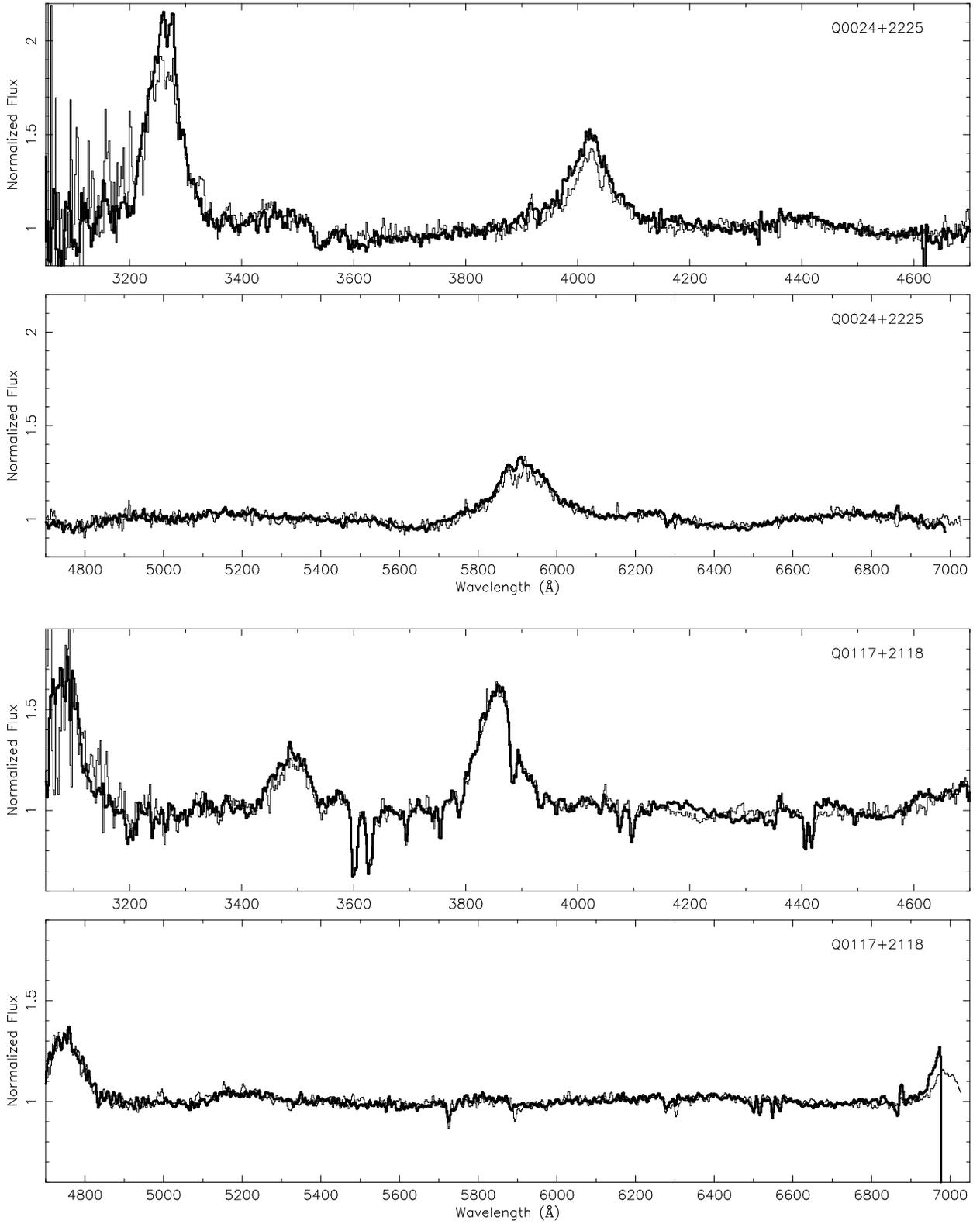}{8 in}{0}{100}{100}{-306}{-90}
\caption[]
{Continuum-flattened spectra of the 30 quasars in our survey.
The blue and red spectra are displayed in separate panels for each object.  
The first epoch data are drawn with the heavier line.}
\label{figures:quasars}
\end{figure}

\newpage
\clearpage

\begin{figure}
\plotfiddle{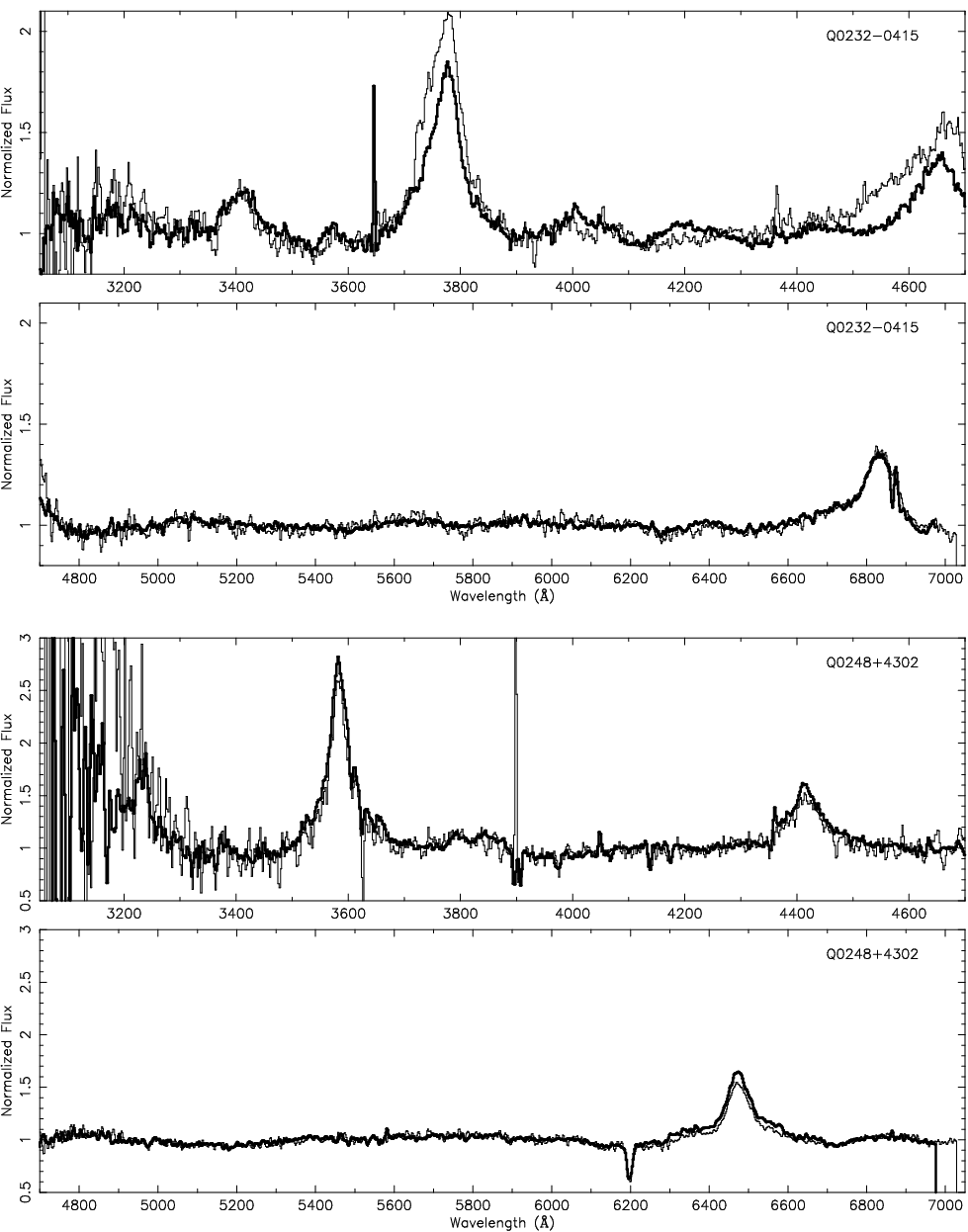}{8 in}{0}{100}{100}{-306}{-90}
\figurenum{1}
\caption[]
{{\it Continued.}}
\end{figure}

\newpage
\clearpage

\begin{figure}
\plotfiddle{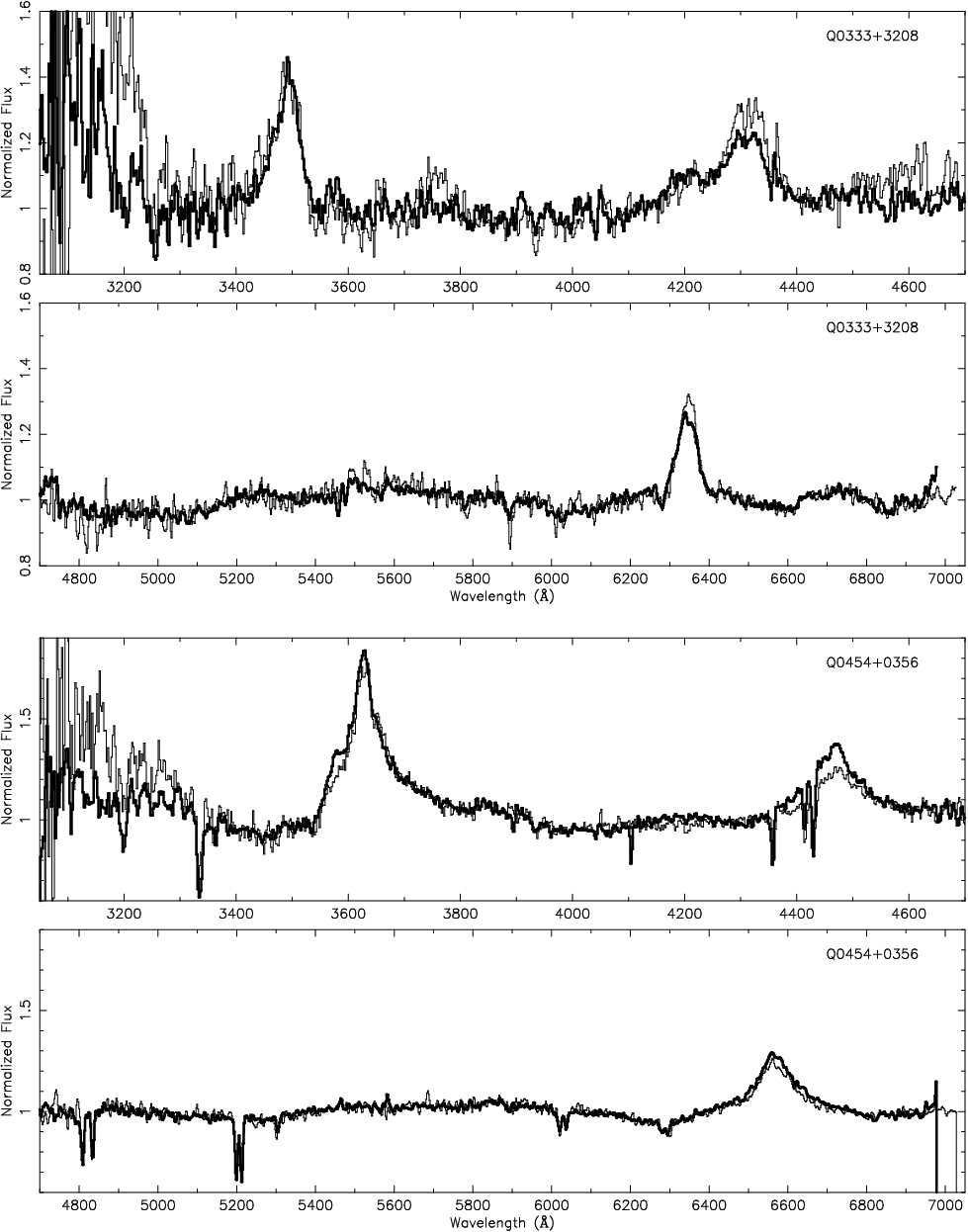}{8 in}{0}{100}{100}{-306}{-90}
\figurenum{1}
\caption[]
{{\it Continued.}}
\end{figure}

\newpage
\clearpage

\begin{figure}
\plotfiddle{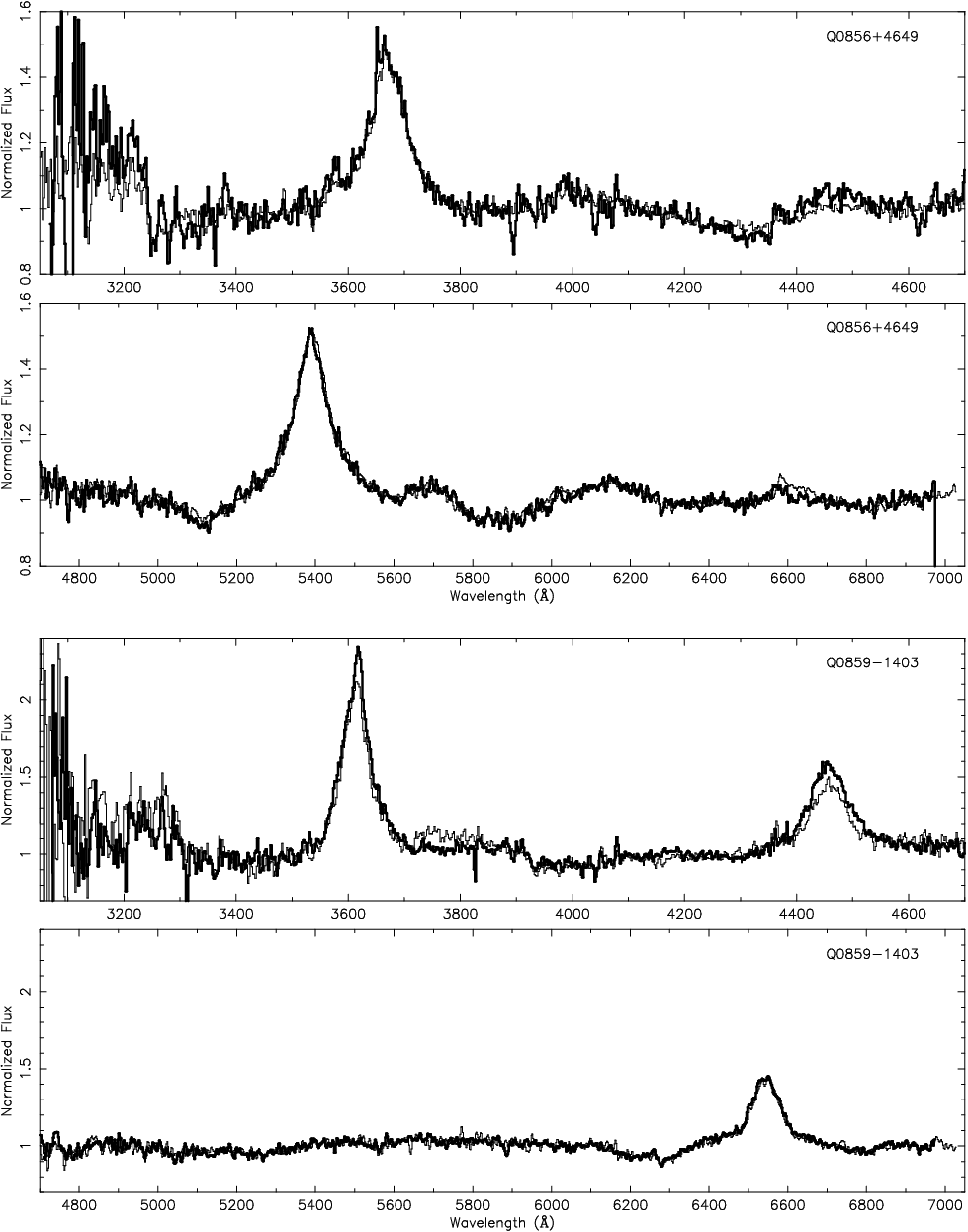}{8 in}{0}{100}{100}{-306}{-90}
\figurenum{1}
\caption[]
{{\it Continued.}}
\end{figure}

\newpage
\clearpage

\begin{figure}
\plotfiddle{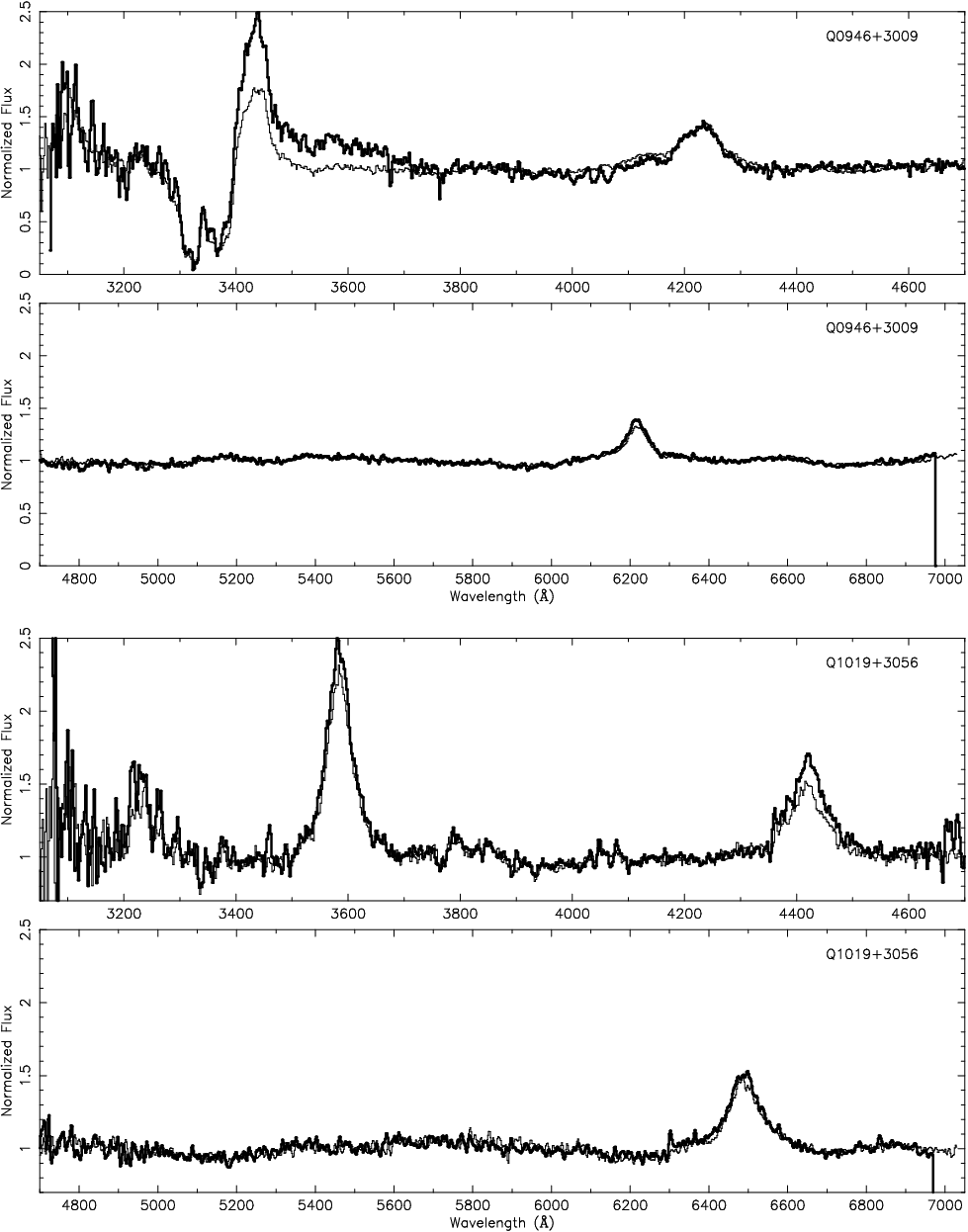}{8 in}{0}{100}{100}{-306}{-90}
\figurenum{1}
\caption[]
{{\it Continued.}}
\end{figure}

\newpage
\clearpage

\begin{figure}
\plotfiddle{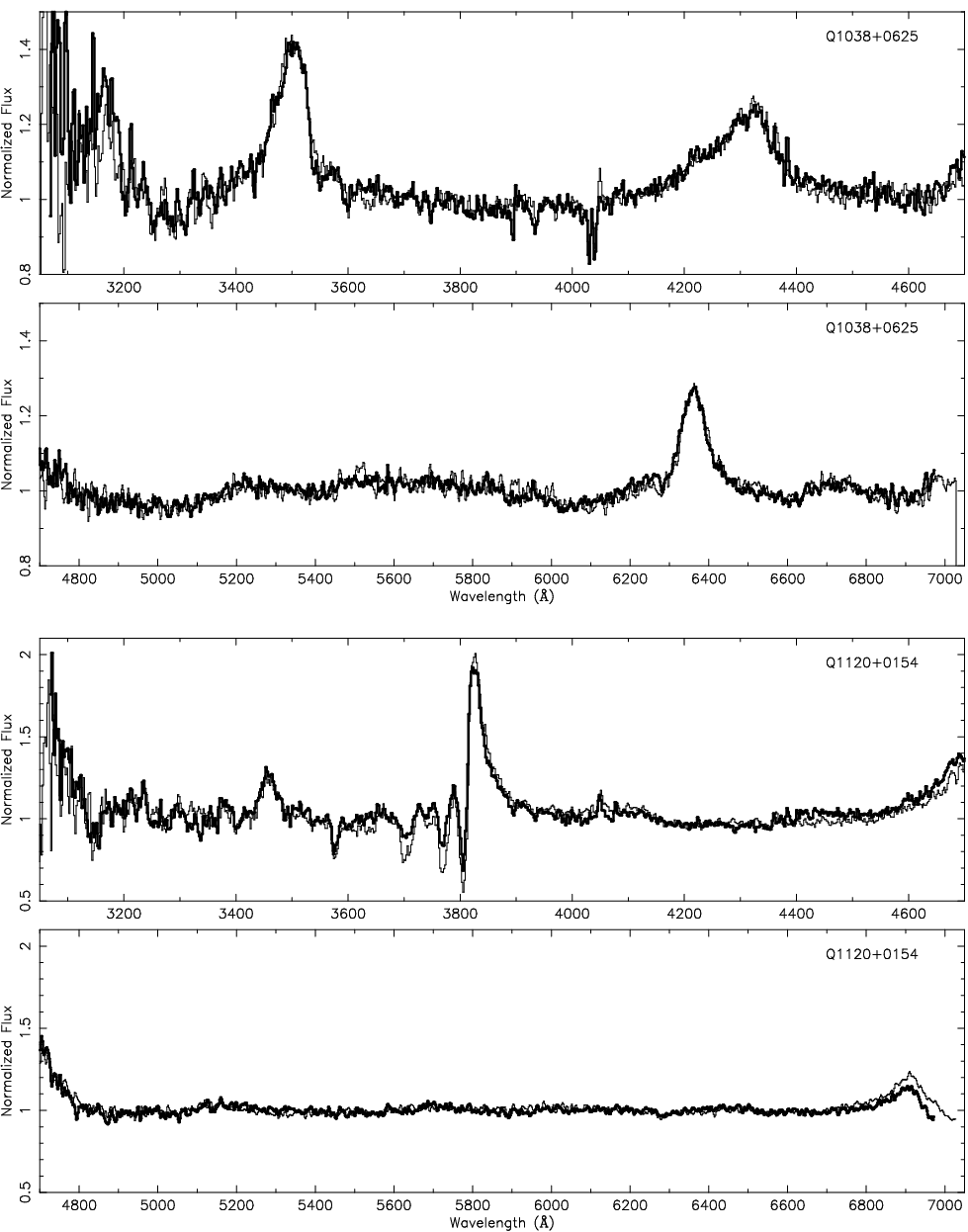}{8 in}{0}{100}{100}{-306}{-90}
\figurenum{1}
\caption[]
{{\it Continued.}}
\end{figure}

\newpage
\clearpage

\begin{figure}
\plotfiddle{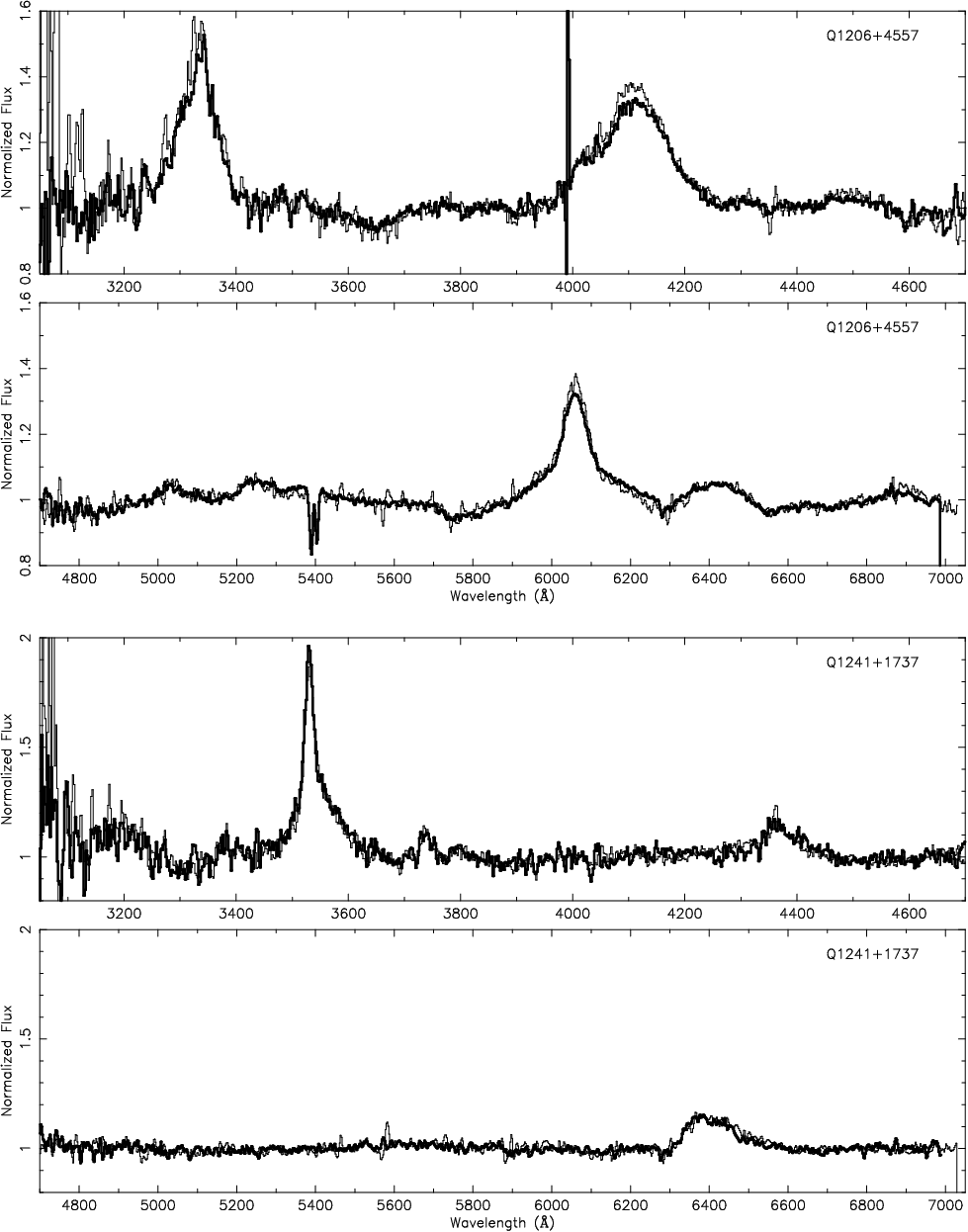}{8 in}{0}{100}{100}{-306}{-90}
\figurenum{1}
\caption[]
{{\it Continued.}}
\end{figure}

\newpage
\clearpage

\begin{figure}
\plotfiddle{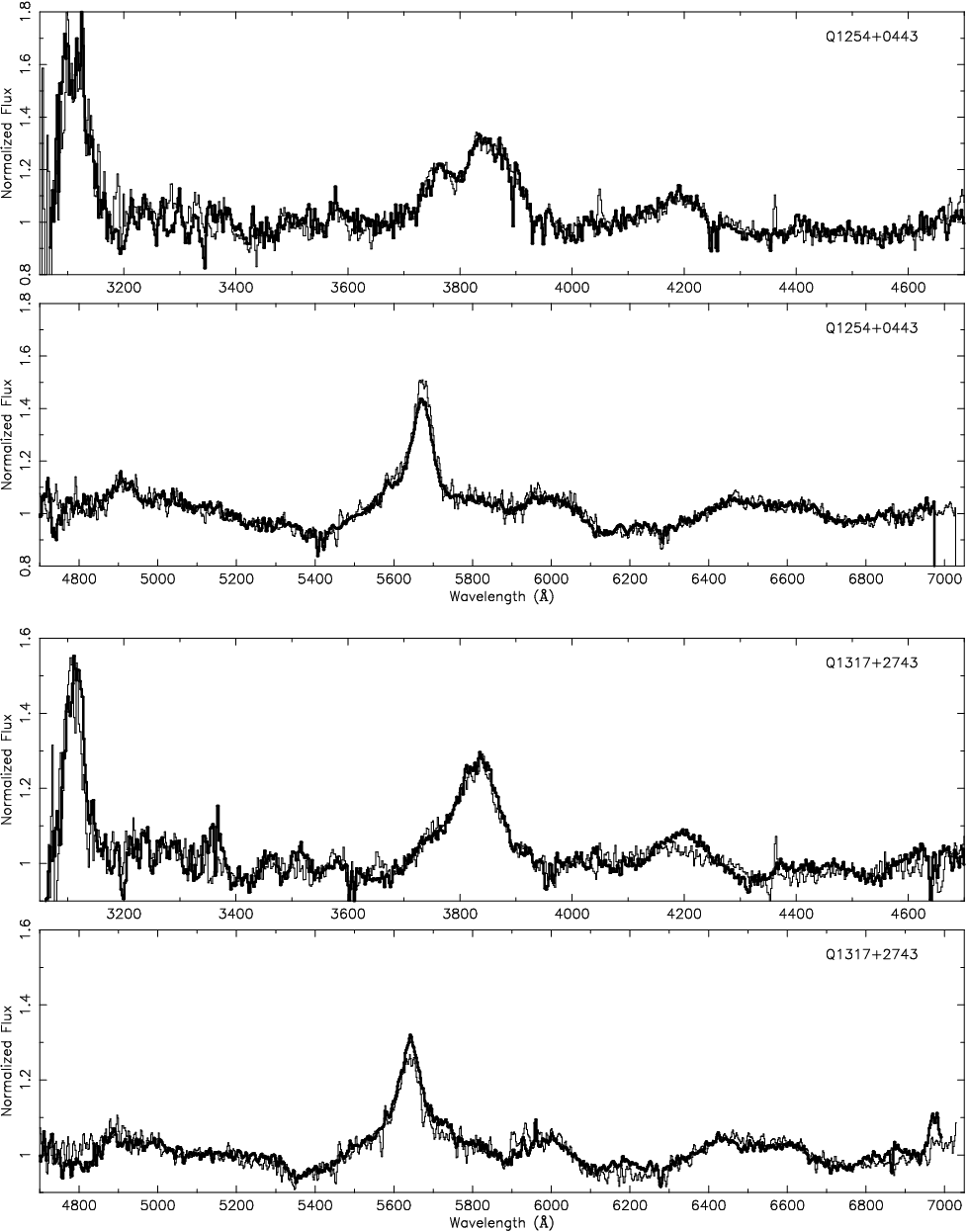}{8 in}{0}{100}{100}{-306}{-90}
\figurenum{1}
\caption[]
{{\it Continued.}}
\end{figure}

\newpage
\clearpage

\begin{figure}
\plotfiddle{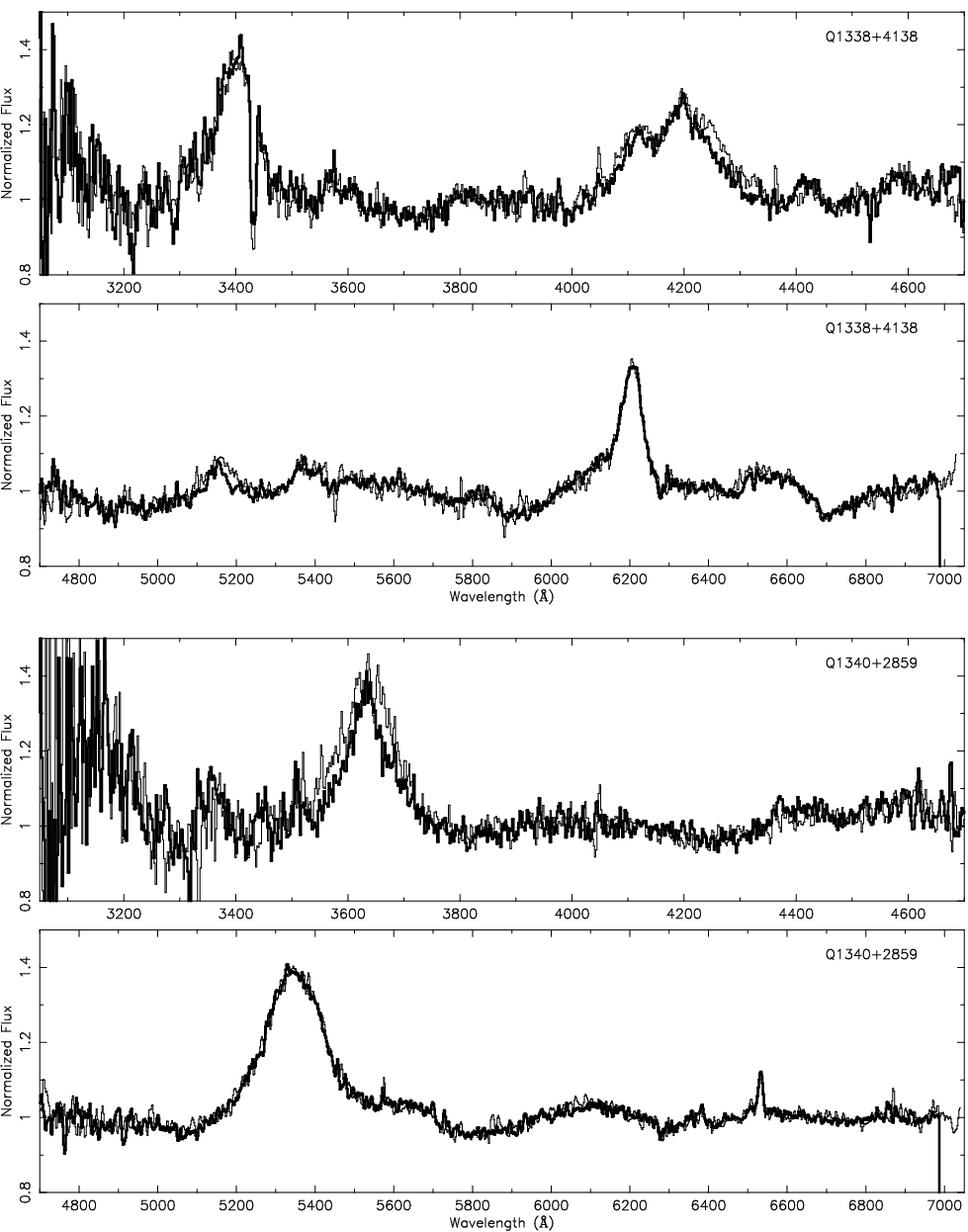}{8 in}{0}{100}{100}{-306}{-90}
\figurenum{1}
\caption[]
{{\it Continued.}}
\end{figure}

\newpage
\clearpage

\begin{figure}
\plotfiddle{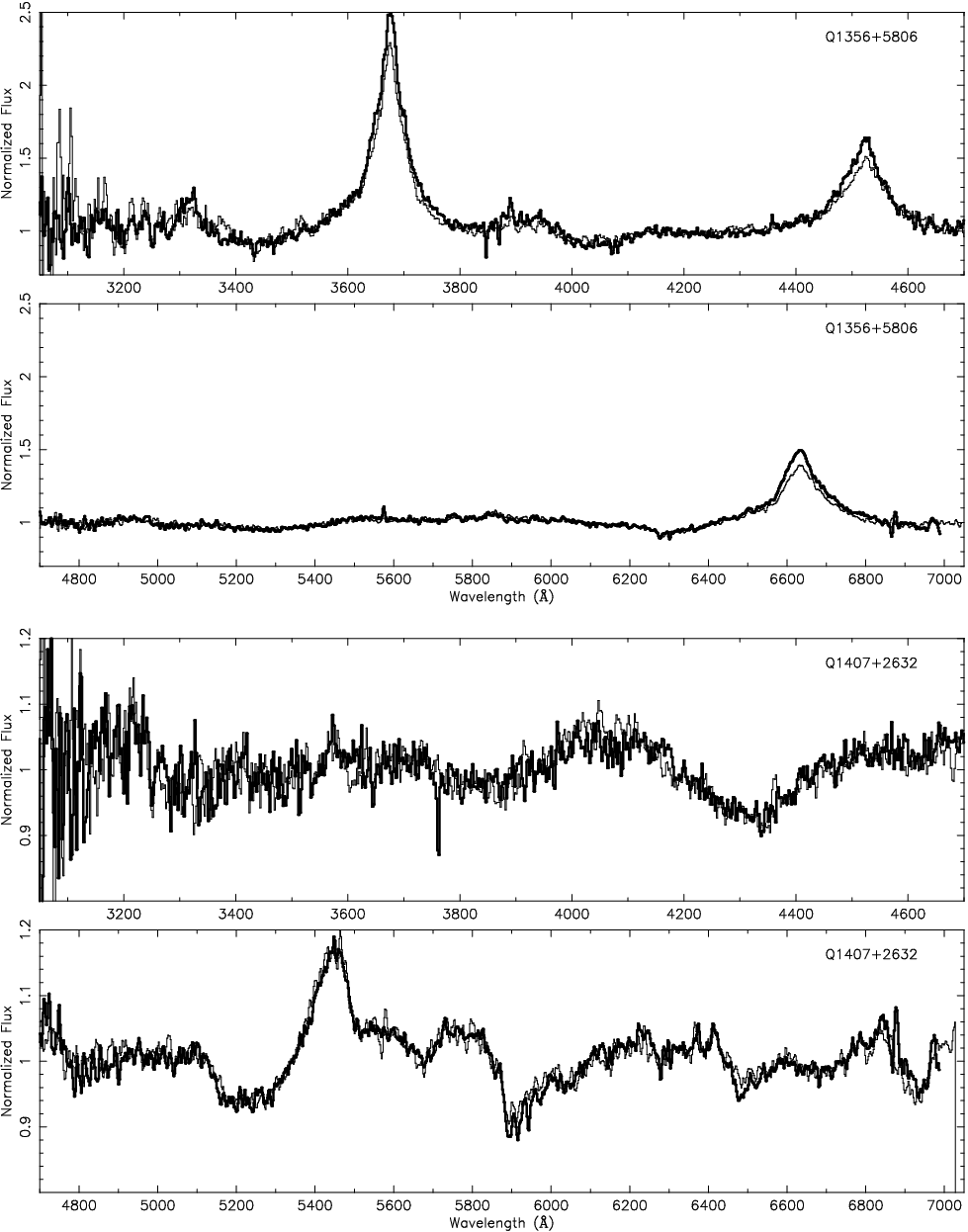}{8 in}{0}{100}{100}{-306}{-90}
\figurenum{1}
\caption[]
{{\it Continued.}}
\end{figure}

\newpage
\clearpage

\begin{figure}
\plotfiddle{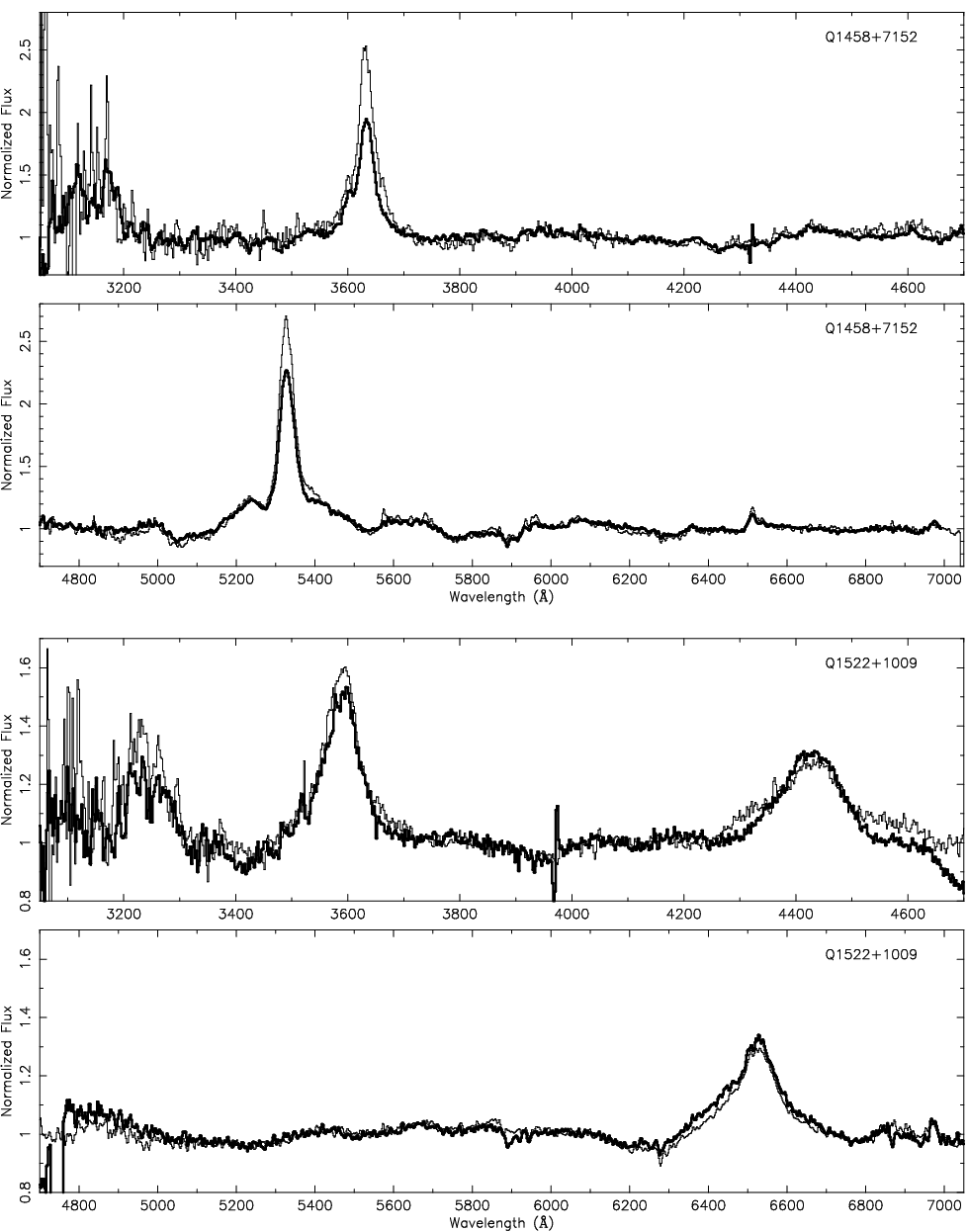}{8 in}{0}{100}{100}{-306}{-90}
\figurenum{1}
\caption[]
{{\it Continued.}}
\end{figure}

\newpage
\clearpage

\begin{figure}
\plotfiddle{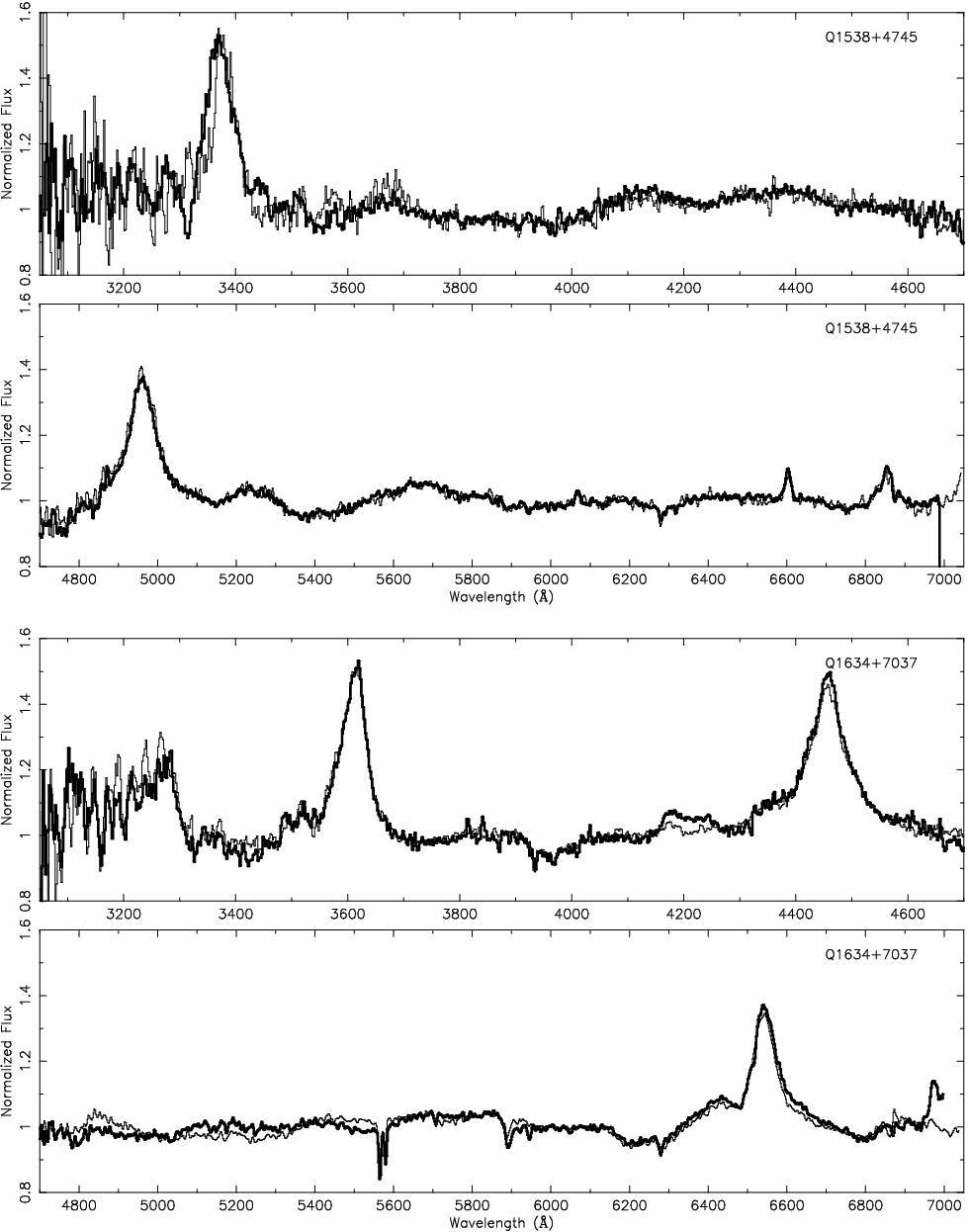}{8 in}{0}{100}{100}{-306}{-90}
\figurenum{1}
\caption[]
{{\it Continued.}}
\end{figure}

\newpage
\clearpage

\begin{figure}
\plotfiddle{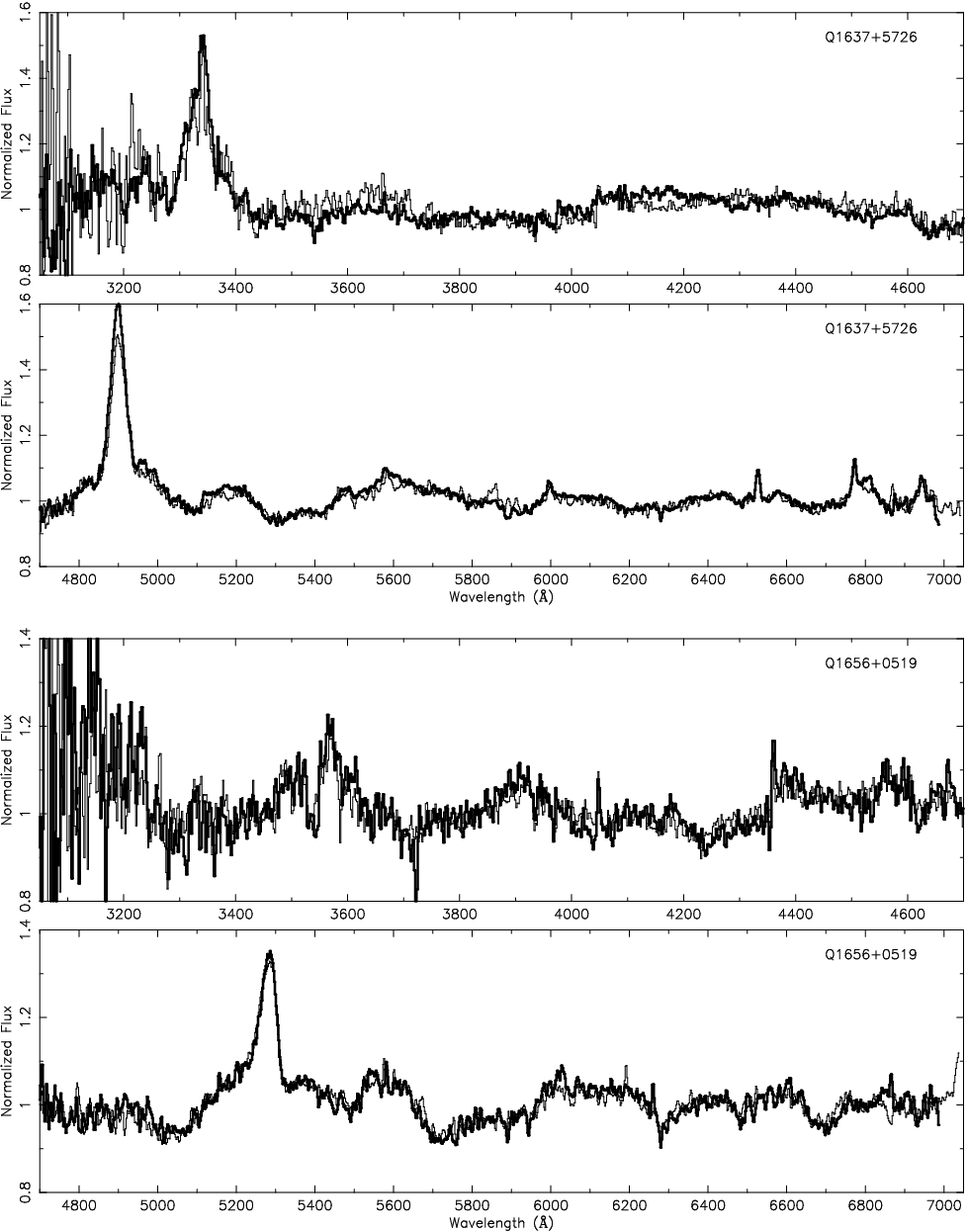}{8 in}{0}{100}{100}{-306}{-90}
\figurenum{1}
\caption[]
{{\it Continued.}}
\end{figure}

\newpage
\clearpage

\begin{figure}
\plotfiddle{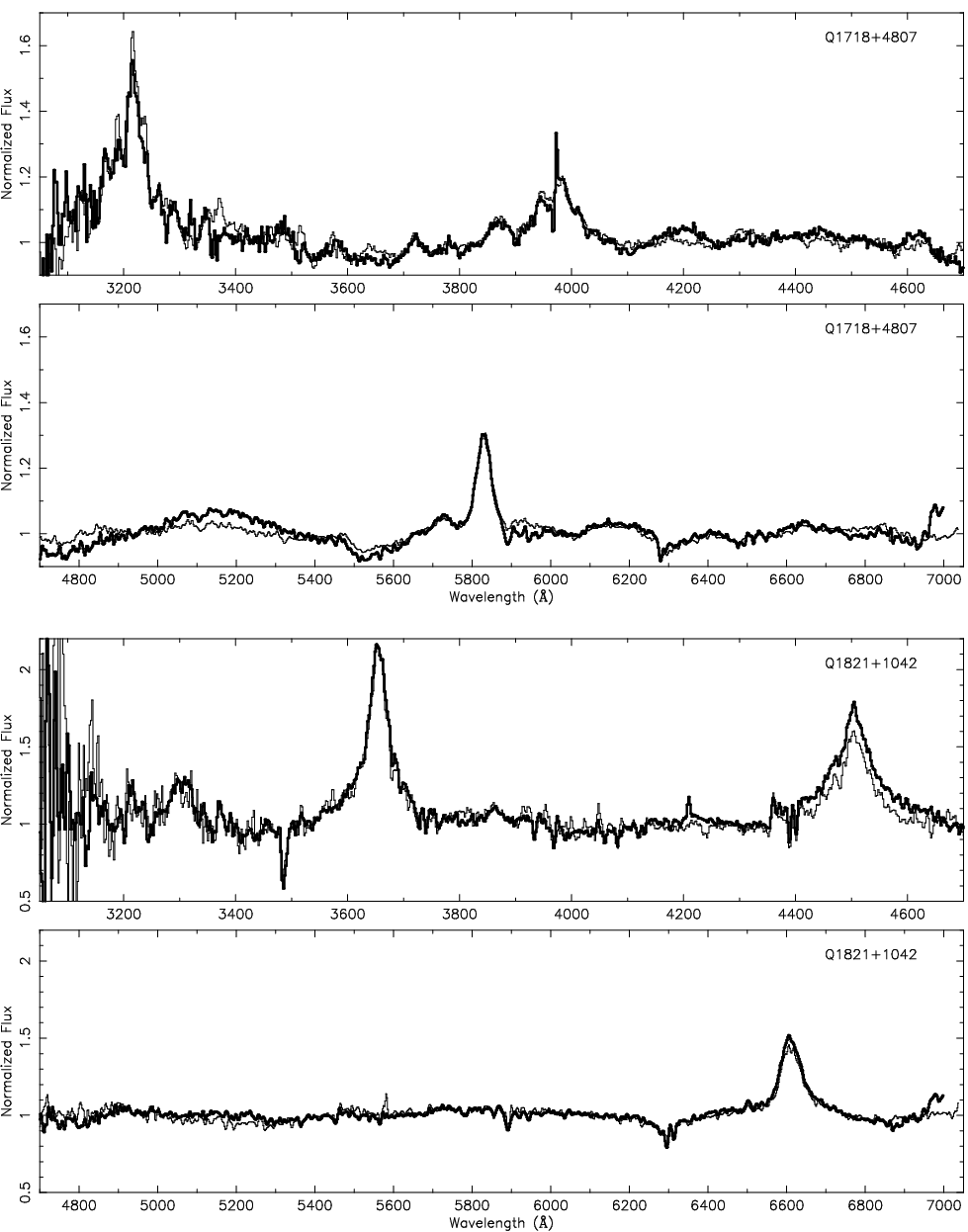}{8 in}{0}{100}{100}{-306}{-90}
\figurenum{1}
\caption[]
{{\it Continued.}}
\end{figure}

\newpage
\clearpage

\begin{figure}
\plotfiddle{qplot_page15.ps}{8 in}{0}{100}{100}{-306}{-90}
\figurenum{1}
\caption[]
{{\it Continued.}}
\end{figure}

\newpage
\clearpage

\newpage
\clearpage

\end{document}